\documentstyle[preprint,aps]{revtex}
\draft
\begin{document}
\title{Effective one--band electron--phonon Hamiltonian for nickel perovskites}
\author{J. Loos$^1$ and H. Fehske$^2$} 
\address{$^1$Institute of Physics, Czech Academy of Sciences, 16200
Prague, Czech Republic\\
$^2$Physikalisches Institut, Universit\"at Bayreuth,
D--95440 Bayreuth, Germany}
\date{Bayreuth, 1 February 1997}
\maketitle
\input{epsf}
\def\gsim{\hbox{$\lower1pt\hbox{$>$}\above-1pt\raise1pt\hbox{$\sim$}$}}
\def\lsim{\hbox{$\lower1pt\hbox{$<$}\above-1pt\raise1pt\hbox{$\sim$}$}}
%\newcommand{\gapro}
%  {\raisebox{-0.25ex} {$\,\stackrel{\scriptscriptstyle>}%
%    {\scriptscriptstyle\sim}\,$}}
%\newcommand{\lapro}
%   {\raisebox{-0.25ex} {$\,\stackrel{\scriptscriptstyle<}%
%    {\scriptscriptstyle\sim}\,$}}
\def\cH{{\cal{H}}}
\begin{abstract}
Inspired  by recent experiments on the Sr--doped nickelates, 
$\rm La_{2-x}Sr_xNiO_4$, we propose a minimal microscopic model 
capable to describe the variety of the observed quasi--static 
charge/lattice modulations and the resulting magnetic and 
electronic--transport anomalies. 
Analyzing the motion of low--spin $(s=1/2)$ holes in a high--spin $(S=1)$
background as well as their their coupling to the in--plane
oxygen phonon modes, we construct a sort of generalized 
Holstein t--J Hamiltonian for the $\rm NiO_2$ planes, which contains
besides the rather complex ``composite--hole'' hopping part 
non--local spin--spin and hole--phonon interaction terms.
\end{abstract}
%\thispagestyle{empty}
%\newpage
\narrowtext
Charge carrier doping of transition metal oxides with perovskite 
related structure induces remarkable phenomena, such as high--temperature
superconductivity in  cuprates (e.g., $\rm La_{2-x}Sr_xCuO_4$), 
intrinsic incommensurate charge and spin ordering in non--metallic nickelates 
$\rm La_{2-x}Sr_xNiO_4$ [LSNO(x)], metal--insulator transition and  
colossal magneto-resistance in Mn--oxides (e.g., $\rm La_{1-x}Ca_xMnO_3$).
All these phenomena are strongly concentration dependent and the experiments
suggest the decisive role of interconnection between the spin-- and charge
correlations and the lattice and transport properties for their 
emergence~\cite{Trea95,Jiea94}.

In this paper we derive an effective Hamiltonian 
describing the interplay of charge--, spin-- and lattice degrees of freedom
in doped Ni--oxides. As revealed by recent neutron scattering of 
LSNO(x)~\cite{TBS96}, the stripe order of both
charge and spin densities in general is found to be incommensurate 
in the low--density region ($x\stackrel{<}{\sim} 0.3$); 
commensurability is restricted to very special 
values of $x$, such as 1/3 and 1/2~\cite{Chea94}.    
The rich variety of charge and spin ordering  
accompanied by the transport anomalies 
in nickelates~\cite{Chea94,Haea92,Raea96} deserves the 
attention not only by itself, but also with respect to understanding the
superconducting state in isostructural cuprates. 
In fact, the incommensurate (stripe--like) 
spin--, charge--correlations and lattice structure modulations
are observed also in the metallic cuprates
but there they are of dynamical and very short--range character~\cite{Trea95}.

The parent compound of the LSNO(x) system is the 
antiferromagnetic (AF) insulator 
$\rm La_2NiO_4$ with a N\'{e}el temperature
$T_N\approx 330$~K and an in--plane exchange constant $J\approx 30$~meV.
The magnetic  $\rm 3d^8$  $\rm Ni^{2+}$ ions, having holes in
$\rm 3d_{x^2-y^2}$,  $\rm 3d_{3z^2-1}$ orbitals, are in the 
high--spin state (HSS) with $S=1$ according to Hund's rule. Doping
this parent compound induces additional holes in the $\rm NiO_2$ plane,
but, in contrast to the superconducting cuprates, the layered 
nickel oxides  LSNO(x) become metallic only near $x\approx 1$.

An additional hole in the $\rm NiO_2$ plane quasi-localized at some 
$\rm Ni^{2+}$ ion aligns its spin antiparallel to the $S=1$ spin 
of the ion due to the strong effective on--site interaction originated  
by the crystal field effect (overcoming the 
Hund's rule coupling)~\cite{ZO93}. 
The resulting low--spin state (LSS) with total spin 1/2 
is tightly bound to the moving charge carrier forming a so--called
``Zhang--Rice doublet''~\cite{Raea96}, which is the counterpart to the usual
Zhang--Rice singlet~\cite{ZR88a} in the cuprates. Aiming at the 
construction of an one--band model, i.e., 
a sort of generalization of the t--J model
for the $\rm NiO_2$ plane, we have to take into account the constraints put
on the motion of the composite--hole LSS by the background of 
correlated HSS of $\rm Ni^{2+}$ ions.

The two configurations, corresponding to an extra hole trapped at one of the 
two nearest neighbor (NN) $\rm Ni^{2+}$ ions of the bond $\langle ij \rangle$  
($i,\;j$ label the sites of the square lattice built up by the Ni ions 
in the a--b plane), are connected  
by an effective transition constant determined by the second order effect
of the intermediate configuration with the hole in the p--orbital of the 
central oxygen ion~\cite{Taea93}. 
Assuming the orbital $\rm d_{jx}^h$ of the hole
localized at the site $j$ to be nearly the same as the orbital 
$\rm d_{x^2-y^2}$ of the Ni--ion coupled to this extra hole 
(see Table~I; in the following the indices $x$,
$z$ stand for the orbitals $\rm d_{x^2-y^2}$, $d_{3z^2-1}$, respectively), 
we shall take orbitals  $\rm d_{jx}^h$, 
$\rm d_{jx}^{}$ to play equivalent roles in the hopping process (and likewise
for analogical orbitals related to the site $i$). This assumption leads to two
possible ways in which the two configurations (differing by the localization
of the extra hole in the bond $\langle ij \rangle$) are connected 
(cf. Table~I).

Moreover, the hopping rate of the hole from $j$ to $i$ also depends on 
the spin states of both configurations leading to different prefactors
in front of the effective transfer constant $t$ (which is determined
by the overlap of the d-- and  p--orbital functions $\propto t_{pd}$). 
According to Serber's method~\cite{Se34} 
(generalizing the Dirac's spin Hamiltonian), the transition 
matrix elements implying the effect of spin states associated with the 
initial and final configurations, are given by the matrix elements of the
sum of the following operators
\begin{eqnarray}
  \label{Htde}
{\cal H}_t^{(d)}&=&t\,{\cal Q}_{j}^{\mbox{\tiny HSS}}{\cal Q}_{i}^{\mbox{
\tiny LSS}}\,{\cal P}_{\mbox{\tiny I}}^s\,
{\cal Q}_{j}^{\mbox{\tiny LSS}}{\cal Q}_{i}^{\mbox{\tiny HSS}}\,,\\
{\cal H}_t^{(e)}&=&
-t\,{\cal Q}_{j}^{\mbox{\tiny HSS}}{\cal Q}_{i}^{\mbox{
\tiny LSS}}\,{\cal P}_{\mbox{\tiny 12}}^s\,
{\cal Q}_{j}^{\mbox{\tiny LSS}}{\cal Q}_{i}^{\mbox{\tiny HSS}}\,,
\end{eqnarray}
acting in the spin function space of two NN Ni ions and one extra hole. 
The identical permutation operator ${\cal P}_{\mbox{\tiny I}}^s$ and the
transposition operator ${\cal P}_{\mbox{\tiny 12}}^s$, both acting on the 
spin variables with indices 1, 2, correspond to the direct--type
and exchange--type hole transfers, respectively, distinguished in Table~I. 
The operators ${\cal Q}_{k}^{\mbox{\tiny LSS}}$ 
(${\cal Q}_{k}^{\mbox{\tiny HSS}}$) project the spin functions
pertaining to site $k$ on the subspace of LSS (HSS). Consequently,
the projection operators restrict the motion of the composite hole 
to the subspace of LSS (for the hole occupied sites) and HSS 
(for the hole unoccupied ones).

The matrix elements of the transitions between  configurations having 
the total spin projections $M_T=1/2$ and  $M_T=3/2$ are given by the 
sum of matrices $[{\cal H}_t^{}] =[{\cal H}_t^{(d)}]+[{\cal H}_t^{(e)}]$ 
as follows
\begin{eqnarray} 
[{\cal H}_t^{(d)}]&=&\!\!\begin{array}{cc}
&|i0, j+\rangle\;\; |i1, j-\rangle \;\; |i1, j+\rangle
\;\; \\[0.2cm]
\begin{array}{c}
\langle i+, j0|\\[0.2cm]
\langle i-, j1|\\[0.2cm]
\langle i+, j1|
\end{array}
&\!\!\!\left[
\begin{array}{ccc}
\quad\frac{1}{3}t\quad\;&\quad\frac{\sqrt{2}}{3}t\quad\;&\quad0\qquad\\[0.2cm] 
\frac{\sqrt{2}}{3}t&0&0\quad\\[0.2cm]
0&0&\frac{2}{3}t\quad
\end{array}\right]\!\!
\end{array}
\end{eqnarray}
\begin{equation} 
[{\cal H}_t^{(e)}]=\frac{1}{2}[{\cal H}_t^{(d)}]\,.
\end{equation}
Here the whole numbers in the bra-- and ket--vectors label the spin
projection belonging to HSS, whereas spin--up and spin--down LSS are
denoted by $+$ and $-$, respectively. The remaining non--zero
matrix elements corresponding to the transitions with the  
$M_T=-1/2$ and  $M_T=-3/2$ conservation are connected with the previous ones
by the time--reversal operation which leaves the matrices (3), (4) unchanged.
Therefore, adding (3) and (4) three types of hopping processes with effective 
transition constants  $t/2$ and $t/\sqrt{2}$, and $t$ are obtained.

Having determined the transition matrix elements of the hopping processes,
the effective transport Hamiltonian in the space of LSS and HSS of
Ni--ions may be written down. To this end we introduce the tensor product  
space $\cH^S \otimes \cH^h$
\begin{eqnarray}
  \label{tps}
    \cH^S&=&\prod_i\otimes \{|im\rangle\}\qquad m=\pm 1,\, 0\,,\\
    \cH^h&=&\prod_i\otimes \{|i\sigma\rangle,|i0\rangle\}\qquad
\sigma=\uparrow, \downarrow\,,
\end{eqnarray}
where $|i,m\rangle$ are the eigenfunctions of spin $S=1$ operators
$\vec{S}^2_i$, $S^z_i$ at site $i$ with spin projection $m$. 
On the other hand, $|i\sigma\rangle$ means an eigenfunction of
of the spin $s=1/2$ operators $\vec{s}^{\;2}_i$, $s^z_i$ of an additional
hole at site $i$ with spin projection up or down. The hole state $|i0\rangle$
corresponds to no extra hole at $i$.

The  HSS creation and annihilation in the hopping 
process will be described by means of the operators~\cite{ZO93}
\begin{eqnarray}
  \label{Bo}
  B^\dagger_{i,1}&=&b^\dagger_{ix,\uparrow}b^\dagger_{iz,\uparrow},\quad
  B^\dagger_{i,-1}=b^\dagger_{ix,\downarrow}b^\dagger_{iz,\downarrow}\\
  B^\dagger_{i,0}&=&
  (b^\dagger_{ix,\uparrow}b^\dagger_{iz,\downarrow}
  +b^\dagger_{ix,\downarrow}b^\dagger_{iz,\uparrow})/\sqrt{2}
\end{eqnarray}
defined by the Schwinger boson creation operators
$b^\dagger_{ix,\sigma}$, $b^\dagger_{iz,\sigma}$.
Denoting by $|0\rangle_B$ the boson vacuum (with respect to
the closed d--shells Ni configuration), the vectors
\begin{equation}
  \label{isms}
   |i,m\rangle=B_{i,m}^\dagger\,|0\rangle_B^{}\;,   
\end{equation} 
and operators
\begin{eqnarray}
  \label{So}
   S_i^+&=&\sqrt{2}\, ( B^\dagger_{i,1}B_{i,0}^{}
   + B^\dagger_{i,0}B_{i,-1}^{})\\
   S_i^z&=&  B^\dagger_{i,1}B_{i,1}^{}-B^\dagger_{i,-1}B^{}_{i,-1}
\end{eqnarray}
constitute a representation of spin $S=1$ operators and states. 
With the above definitions of $B$--operators, the following 
relations become evident:
\begin{equation}
\label{bimo}
B_{i,m}|0\rangle_B^{}=0,\quad 
B_{i,m}|jm^\prime\rangle=\delta_{ij}\delta_{mm^\prime}|0\rangle_B^{}\,.
\end{equation}
Accordingly, the HSS related to site $i$ can be represented by the vectors
\begin{equation}
  \label{isms2}
   |im\rangle|i0\rangle =B_{i,m}^\dagger\,|0\rangle_B^{}|i0\rangle\;.   
\end{equation}

Now let us consider the LSS of composite holes formed by an extra hole
antiferromagnetically tightly bound to the spin $S=1$ of the $\rm Ni^{2+}$
ion. Defining the Hubbard operators
\begin{equation}
\label{huo} 
X_{i}^{\sigma 0}=|i\sigma\rangle\langle i0|
\end{equation}
in the single--occupation space ${\cal H}^h$ and taking into account the 
interior structure of the composite holes (which was neglected in the
work of Zaanen and Ole\'{s}~\cite{ZO93}), the LSS will 
be expressed by means of the Clebsch--Gordon coefficients as follows:
\begin{eqnarray}  
\label{Lss} 
|i+\rangle\! &=&\!
\mbox{\small $\frac{1}{\sqrt{3}}$}
(- B_{i,0}^\dagger X_i^{\uparrow 0} + \mbox{$\small\sqrt{2}$}
 B_{i,1}^\dagger X_i^{\downarrow 0})\,|0\rangle_B^{}|i0\rangle\\
|i-\rangle \!&=&\!
\mbox{\small $\frac{1}{\sqrt{3}}$}
(\mbox{$\small - \sqrt{2} $}B_{i,-1}^\dagger X_i^{\uparrow 0} +
 B_{i,0}^\dagger X_i^{\downarrow 0})\,|0\rangle_B^{}|i0\rangle\,.
\end{eqnarray}

Then, using the transition matrix elements given by (1)--(4) 
as well as the representation of HSS and LSS by (13), (15), (16), the 
effective transport Hamiltonian ${\cal H}_t$ may be rewritten as
\begin{equation}
  \label{ht1}
  \cH_t=t\sum_{\langle i,j\rangle}(C^\dagger_{i,\mbox{\small $\frac{1}{2}$}}
C_{j,\mbox{\small $\frac{1}{2}$}}^{}
+C^\dagger_{i,\mbox{-\small $\frac{1}{2}$} }
C_{j,\mbox{-\small $\frac{1}{2}$} }^{})\,,
\end{equation}
with
\begin{equation}
\label{Co1}
C_{j,\mbox{\small $\frac{1}{2}$}}^{}=\bar{B}_{j,0}^\dagger \left[
\mbox{\small $-\frac{1}{\sqrt{3}}$}
 B_{j,0}^{} X_j^{0 \uparrow}+\mbox{\small $\sqrt{\frac{2}{3}}$}
  B_{j,1}^{} X_j^{0 \downarrow}\right] 
+B_{j,-1}^\dagger \left[\mbox{\small $-\sqrt{\frac{2}{3}}$}
 B_{j,-1}^{} X_j^{0 \uparrow}+\mbox{\small $\frac{1}{\sqrt{3}}$}
  B_{j,0}^{} X_j^{0 \downarrow}\right]\,,
\end{equation}
where $\bar{B}^\dagger_{j,0}=\mbox{\small $\frac{1}{\sqrt{2}}$} 
B^\dagger_{j,0}$. The first term on the right hand side of (17) describes 
the four hole transport processes which are connected with the 
transfer of the spin projection equal to 1/2. The second term in (17),
comprising the remaining four hopping processes obtained from the previous
ones by the time reversal operation, corresponds to the transfer of spin 
projection equal to (--1/2). 

Obviously, in $\cH_t$ the Hubbard operators of holes are 
coupled to the Schwinger boson operators,  
showing thus a rather complicated dependence of the hole transport 
on the spin background. We can even go a step further by expressing the  hole
Hubbard operators in terms of decoupled charge (holon) and spin variables.
Using a slightly modified treatment proposed recently for the mapping 
of the t--J model onto the tensor product space of holon and
spin--1/2 states~\cite{Lo96}, the Hubbard operators~(14) are 
obtained  in terms of independent holon 
$(h_i)$ and pseudo-spin--1/2 ($\tilde{s}_i$) operators
as
\begin{eqnarray}
  \label{Xo}
  X_i^{0\uparrow}&=&h_i^{}(\tilde{s}_i^+\tilde{s}_i^-
  +\mbox{e}^{i\varphi}\tilde{s}_i^-)/\sqrt{2}\,,\\
  X_i^{0\downarrow}&=&h_i^{}(\tilde{s}_i^++\mbox{e}^{i\varphi}\tilde{s}_i^-
  \tilde{s}_i^+)/\sqrt{2}\,,
\end{eqnarray}
where the arbitrary phase factor $\varphi$ has no effect on the 
matrix elements of physical variables. The local number operator 
of spinless fermionic holons,
\begin{equation}
  \label{no}
 h^\dagger_ih^{}_i= \sum_\sigma X_i^{\sigma 0}X_i^{0\sigma}\,,
\end{equation}
has eigenvalues $n_i^h=0,\,1$, and the spin operators of the physical
hole--spins are connected with the pseudo-spin operators by
\begin{equation}
  \label{so}
s_i^{\pm}=h^\dagger_ih^{}_i\tilde{s}_i^{\pm}\,,
\;\;\;s_i^{z}=h^\dagger_ih^{}_i\tilde{s}_i^z\,.
\end{equation}
The corresponding representation of the operators 
$C_{j,\mbox{\small $\pm\frac{1}{2}$}}$ defined by (18) is given by 
\begin{eqnarray}
\label{Co2}
C_{j,\mbox{\small $\frac{1}{2}$} }^{} &=&\mbox{\small $\frac{1}{\sqrt{6}}$} h_j\left[\left(
\mbox{\small $\sqrt{2}$}
\bar{B}_{j,0}^\dagger B_{j,1}^{} +B_{j,-1}^\dagger B_{j,0}^{}\right)
\left(\tilde{s}_j^++\mbox{e}^{i\varphi}\tilde{s}_j^-
  \tilde{s}_j^+\right)
\right.\nonumber\\
&&\;\; -\left.
\left(\bar{B}_{j,0}^\dagger B_{j,0}^{} 
+\mbox{\small $ \sqrt{2}$} 
B_{j,-1}^\dagger B_{j,-1}^{}\right)\left(\tilde{s}_j^+\tilde{s}_j^-
  +\mbox{e}^{i\varphi}\tilde{s}_j^-\right)\right],\nonumber\\
\end{eqnarray}
and the effective transport Hamiltonian~(17) becomes 
\begin{equation}
  \label{ht2}
   \cH_t=\sum_{\langle i,j\rangle} \hat{t}_{ij}^{} h^\dagger_i h^{}_j\,.
\end{equation}
The latter expression (24) of $\cH_t$  
has the form of a holon hopping Hamiltonian with the hopping constant
$t$ replaced by the transfer operator $\hat{t}_{ij}$ 
which depends on the degrees of freedom of the spin--1 background,
as well as on the spin variables of the charge carriers.

To demonstrate more explicitly the dependence 
of the charge transport on the spin background 
we shall consider the strong AF correlations of the $\rm Ni^{2+}$ spins at low
hole concentrations, what enables us to use the spin wave approximation (SWA).
In the spirit of SWA the states with double local spin deviations from the 
N\'{e}el ordering will be disregarded. Accordingly, separating the 
lattice sites into two AF sublattices $\{k\}$, $\{l\}$ characterized by
$S_k^z=S$, $S_l^z=-S$, respectively, the states $|k,-1\rangle$, $|l,1\rangle$ 
of $\rm Ni^{2+}$ ions will be excluded. 
Then the representation of spin operators (10), (11) may be approximated by 
\begin{eqnarray}
  \label{swa}
  S_k^+&\simeq& \sqrt{2} B^\dagger_{k,1}B_{k,0}^{}\,,\;\;\;  
S_k^z \simeq  B^\dagger_{k,1}B_{k,1}^{}\,,\\
  S_l^+&\simeq&\sqrt{2} B^\dagger_{l,0}B_{l,-1}^{},\;\;\; 
S_l^z\simeq -B^\dagger_{l,-1}B_{l,-1}^{}\,, 
 \end{eqnarray}
what makes possible to express the transport Hamiltonian given by (17), (18) 
by means of $S^{\pm}$ (spin wave) operators, up to quadratic terms, as follows:
\begin{eqnarray}
  \label{ht3}
\cH_t^{SWA}&=&\frac{t}{6}\sum_{\langle kl \rangle} \left[\left(
S_k^-S_k^++S_l^-S_k^+\right)X_l^{\uparrow0}X_k^{0\uparrow}
\right.\nonumber\\
&&\qquad\quad 
+\left(
S_l^+S_l^-+S_l^+S_k^-\right)X_l^{\downarrow0}X_k^{0\downarrow}
%\nonumber\\
% &&\qquad \quad
\left.
-2\left(S_k^-+S_l^-\right)X_l^{\uparrow0}X_k^{0\downarrow}
+\mbox{H.c.}\right]\,.
\end{eqnarray}

The difference of (27) with respect to the transport term of the standard
t--J model consists in the spin dependence of the hole transport, 
as well as in the  existence of hopping accompanied by hole--spin flipping. 
A closer resemblance of the t--J model is obtained 
if the transport Hamiltonian (27) is averaged 
with respect to the ground state of AF magnons. 
Following the procedure outlined for the t--J model 
in Ref.~\cite{Lo96}, we get the mean field Hamiltonian  
\begin{equation}
  \label{htswa}
\overline{\cH}^{SWA}_t= t_{eff}^{}\sum_{\langle i,j\rangle} 
h_i^\dagger h_j^{} (g_{ij}^{}+\vec{\tilde{s}}_i\vec{\tilde{s}}_j +
\mbox{\small$\frac{1}{4}$})
\end{equation}
with
\begin{eqnarray}  
\label{teff}
  t_{eff}&=&\frac{t}{3}\left(\langle \delta S^z_k\rangle + 
\mbox{\small $\frac{1}{2}$}
\langle S_k^+ S_l^- \rangle \right)\,,\\
\label{gij}
g_{ij}&=&\left\{\mbox{e}^{i\varphi}
\left[\left(\mbox{\small$\frac{1}{2}$}-\tilde{s}_j^z\right)
\tilde{s}_i^-+\tilde{s}_j^-\left(\mbox{\small$\frac{1}{2}$}
+\tilde{s}_i^z\right)\right]\right.
%\nonumber\\&&
+\mbox{e}^{-i\varphi}
\left.\left[\left(\mbox{\small$\frac{1}{2}$}+\tilde{s}_j^z\right)
\tilde{s}_i^++\tilde{s}_j^+\left(\mbox{\small$\frac{1}{2}$}
-\tilde{s}_i^z\right)\right]\right\}/2\,.
\end{eqnarray}
Here $\langle \delta S_k^z \rangle$ is the reduction of the local
$|S_i^z|$ from the classical value $S$ in the ground state of AF magnons.
The expectation values in (29) vanish in the classical AF N\'{e}el ground 
state. Supposing the quantum antiferromagnet, $t_{eff}$ is non--zero 
owing to the zero--point fluctuations, but it is considerably reduced 
compared with the bare transfer integral.
In this way, the spin correlations lead to strong magnetic 
confinement effects and therefore suppress the charge transport.

In a next step we consider the spin interactions. 
Here we have to take into account the strong
effective on--site interaction, 
leading to the formation of LSS at the hole
occupied sites and the (much weaker) 
superexchange interaction of the NN spins. Thus we get 
\begin{equation}
\label{hsp}
\cH_{ {\cal J}}= {\cal J}_0^{} \sum_i h_i^\dagger h_i^{} 
\vec{\tilde{s}}_i \vec{S}_i+
\sum_{\langle i,j\rangle}  {\cal J}_{ij} \vec{J}_i\vec{J}_j\,,
\end{equation}
where
\begin{equation}
\label{J}
\vec{J}_i=(1- h_i^\dagger h_i^{} )
\vec{S}_i+h_i^\dagger h_i^{}  (\vec{S}_i+\vec{\tilde{s}}_i)\,.
\end{equation} 
The effective on--site coupling constant  ${\cal J}_0$ is given by
the energy difference between the $J=1/2$ (LSS) and $J=3/2$ states
of the $\rm Ni^{2+}$ ion occupied by an additional hole. 
Dagotto's estimate~\cite{DRSM96} of the latter difference 
is $\sim 1.3$~eV, so that the on--site
interaction is much stronger than than all the inter--site ones. 
Note that the superexchange interaction ${\cal J}(n_i^h,n_j^h)$ 
between the total spins $\vec{J}_i$ of the NN cations, 
represented by the second term in~(31), 
depends strongly on the electronic configuration of the bond 
$\langle ij \rangle$~\cite{An50}.  
For pairs of $\rm Ni^{2+}$ ions, ${\cal J}(0,0)$ 
is the superexchange coupling constant in the parent compounds.
Most notably, recent experiments~\cite{TBS96} show that the intersite spin 
interactions play a secondary role in formation the charge--spin stripe 
structure as the charge ordering occurs always at higher temperatures as 
the ordering of spins. 

On the other hand, in the nickelates there is both 
experimental~\cite{CCC93,BE93} and theoretical~\cite{Piea88,ZL94} 
evidence for a strong coupling of the doped holes to the in--plane 
oxygen phonon modes. It is believed that the observed charge and spin 
modulations are driven by a charge segregation in stripe--like structures,
i.e. by phase separation on a mesoscopic length scale, connected with 
breathing--mode polaron formation. The mechanism for such polaron ordering is 
expected to be based on: (i) the non--local character of the electron--phonon
interaction determining the energy gain caused by the bond 
deformation to be proportional to the hole 
population difference of the NN Ni--sites, and 
(ii) on the Coulomb polaron polaron interaction. 
Once the ordering of polarons is established, the distribution of 
LSS and HSS in the lattice is fixed and at sufficiently low temperature 
the stripe spin--ordering arises.

To include the in--plane electron--phonon coupling effects
in our model Hamiltonian, we assume a Holstein--type 
interaction which takes the form
\begin{equation}
\label{hep}
\cH_{h-p}=-A\sum_{\langle i,j\rangle} \hat{u}_{ij}\,
( h_i^\dagger h_i^{}  -  h_j^\dagger h_j^{} )
\end{equation}
for the case considered. Here $A$ denotes the local hole--lattice
coupling constant and  $\hat{u}_{ij}$ is the displacement operator of 
the oxygen in the bond $\langle i j\rangle$.
That is, in our effective single--band description the formation of
polarons is related to a quasi--static (oxygen--nickel) bond deformation
given by  $\langle \hat{u}_{ij}\rangle \neq 0$. As a second order effect 
the overlap integrals $t_{pd}$ and thus the  superexchange 
interactions are affected by a finite  $\langle \hat{u}_{ij}\rangle$  
as well, i.e. we have  
$ {\cal J}_{ij}= {\cal J}(n_i^h,n_j^h,\langle \hat{u}_{ij}\rangle)$.

To this end, as an effective Hamiltonian for the theoretical description 
of the layered nickel perovskites we propose 
the following generalized Holstein t--J model 
\begin{equation}
\label{htjm}
\cH= \cH_t+\cH_{{\cal J}}+\cH_{h-p}+\cH_p\,,
\end{equation}
where $\cH_p$ refers to the bare phonon part given in harmonic approximation.

In summary, we have derived the effective hopping transport Hamiltonian
of spin--1/2 composite holes in the background of spin--1 
$\rm Ni^{2+}$ ions using Serber's results for 
the transition matrix elements between different 
electron configurations. The suppression of the spin--dependent transport
was explicitly demonstrated for the case that the  $\rm Ni^{2+}$--spins 
are strongly AF correlated. The experimentally observed stripe--structures  
appear to be comprehensible on the basis of the strong coupling of doped 
holes to the in--plane oxygen lattice modes, incorporated in the 
proposed effective Hamiltonian by a non--local Holstein--type interaction. 

Finally, we would like to emphasize that applying the approach
described above to the electronic transport in the
ferromagnetic manganese compounds , e.g.
$\rm La_{1-x}Ca_xMnO_3$~\cite{AH55}, shows the two essential differences 
between the doped Ni-- and Mn--oxides. First, 
the itinerant electron of $\rm Mn^{3+}$ couples to the three
$t_{2g}$ electrons according to Hund's rule forming a {\it high--spin state}
with total $S=2$. In contrast the crystal field splitting 
is dominant for $\rm Ni^{3+}$ ions leading 
to a {\it low--spin state} with total spin~1/2.
Secondly, whereas the $\rm e_g$ electrons of Mn--ions are separated 
from the spin background, determined by the localized $t_{2g}$ electrons,
the holes in the $\rm d_{x^2-y^2}$ orbitals of Ni--ions take part in
the hopping, as well as in the formation of spin background.

This work was in part supported by the Grant Agency of
%Czech Republic, Project No. 202/96/0864, and by the Deutsche 
%Forschungsgemeinschaft through SFB~279. 
\begin{table}
\caption{Hole transfer processes.}
\label{tab1}
\begin{tabular}{cccccccc}
\rule[-3mm]{0mm}{7mm}configuration& $\rm d_{jx}^h$ & $\rm d_{jx}^{}$ &  
$\rm d_{jz}^{}$ & p &
$\rm d_{ix}^h$ & $\rm d_{ix}^{}$ &  $\rm d_{iz}^{}$\\\tableline
\rule[-2mm]{0mm}{6mm}initial&1&2&3&&&4&5\\
\tableline
\rule[-2mm]{0mm}{6mm}&\multicolumn{7}{c}{direct--type transfer (d)}\\
intermediate&&2&3&1&&4&5\\
final&&2&3&&1&4&5\\[0.2ex]
\tableline
\rule[-2mm]{0mm}{6mm}&\multicolumn{7}{c}{exchange--type transfer (e)}\\
intermediate&1&&3&2&&4&5\\
final&&1&3&&2&4&5\\
\end{tabular}
\end{table}
\def\baselinestretch{0.95}
%\acknowledgements{This work was in part supported by the Grant Agency of
%Czech Republic, Project No. 202/96/0864, and by the Deutsche 
%Forschungsgemeinschaft through SFB~279.
\bibliography{ref}
\bibliographystyle{phys}
\end{document}